\documentclass[%
 reprint,
 amsmath,amssymb,
 aps,
]{revtex4-2}
\usepackage{graphicx}
\usepackage{dcolumn}
\usepackage{bm}
\usepackage{hyperref}
\DeclareUnicodeCharacter{2212}{-}

\begin{document}

\preprint{APS/123-QED}

\title{Detection of radiatively open systems using an optical cavity }

\author{V. I. Gokul${}^1$, Arun Bahuleyan${}^1$, Raghuveer Singh Yadav${}^1$,  S. P. Dinesh${}^1$, V. R. Thakar${}^1$,\\ Rahul Sawant${}^2$ \& S. A. Rangwala${}^1$}
\affiliation{%
 ${}^1$Raman Research Institute, C.V. Raman Avenue, Sadashivanagar, Bangalore 560080, India \\
 ${}^2$ I-Hub Quantum Technology Foundation, Indian Institute of Science Education and Research Pune, Dr. Homi Bhabha Road, Pashan, Pune 411008 India.
}%

\date{\today}

\begin{abstract}
We experimentally demonstrate a cavity-based detection scheme for a cold atomic ensemble with a radiatively open transition. Our method exploits the collective strong coupling of atoms to the cavity mode, which results in off-resonant probing of the atomic ensemble, leading to a dramatic reduction in losses from the detection cycle. We then show the viability of this frequency measurement for detecting a small number of atoms and molecules by theoretical modelling. Compared with the most commonly used fluorescence method, we show that the cavity-based scheme allows rapid and prolonged detection of the system's evolution with minimal destruction.  

\end{abstract}

\maketitle

\section{Introduction}
While atomic and molecular systems are multi-level systems, their efficient optical detection depends on probing a relatively isolated two-level sub-system~\cite{mot1,bec1,bec2,quantum-jump1}. For most of these quasi two-level systems, fluorescence detection is the method of choice~\cite{mot1,fluorescence1,fluorescence2,fluorescence3,single-atom1,fluorescence5,single-atom2,single-atom3}. However, direct fluorescence imaging is experimentally challenging for the large class of multi-level atoms/molecules~\cite{molecule-imaging1,molecule-imaging2,molecule-imaging3} because of the lack of such isolated sub-systems. Detecting atoms with an open transition, where the decay from the excited state can shelve atoms to multiple dark states, requires alternate detection schemes. Most current detection schemes for cold molecules are destructive and rely on converting molecules back to atoms~\cite{free-atom1,atom2,atom3,atom4,bec2}, ionization of molecules~\cite{rempi1,rempi2,rempi3}, or direct absorption on a high optical density ensemble~\cite{Ni-molecule}. In an article by Sawant \textit{et al.} ~\cite{rahul-molecule}, a non-destructive detection scheme that uses an optical Fabry-Perot (FP) cavity was theoretically discussed for molecules with multiple decay channels; this scheme exploits the collective strong coupling of molecules to the cavity mode.

In this article, we experimentally demonstrate the detection of an open transition in an atomic system (Fig.~\ref{fig1} (a)) using an FP cavity and study the transient state evolution of the atom-cavity system. This helps us mimic the open transitions in molecules and demonstrate the viability of the scheme for them. We then explain the result with a theoretical model and discuss the possibility of generalized detection of molecules using a cavity in this framework. Our method of cavity detection prolongs the atomic population in the resonant two-level sub-system, allowing frequent, precise, low-noise detection of the atoms.

\section{Measuring open transition using a cavity}
\begin{figure}[!t]
	\centering
	\includegraphics[width=0.45\textwidth]{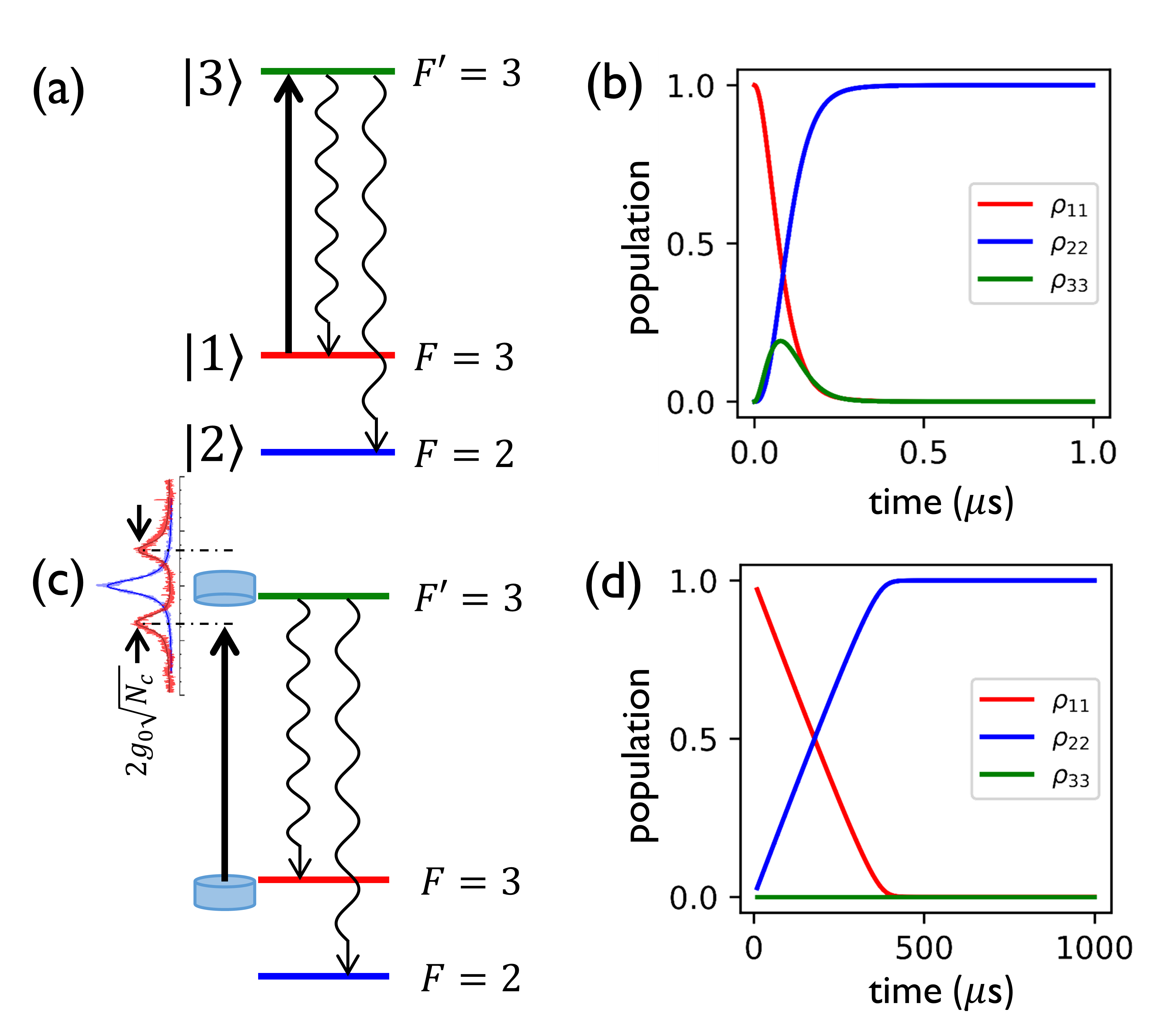}
	\caption{(a) Energy level diagram for the multi-level system with population losses via an open transition. The probe laser addresses the $F=3\leftrightarrow F'=3$ transition in ${}^{85}$Rb. The excited state $F'=3$ can decay to ground states $F=3$ and $F=2$ via dipole-allowed transition. (b) Numerically solved evolution of populations in (a). (c) Schematic of cavity detection of atoms with population loss. (d) Numerically solved evolution of populations in (c). All numerical calculations are done with a probe power of $2$ $\mu$W and diameter $\sim 2$ mm.}
	\label{fig1}
\end{figure}

\begin{figure*}[!t]
	\centering
	\includegraphics[width=0.85\textwidth]{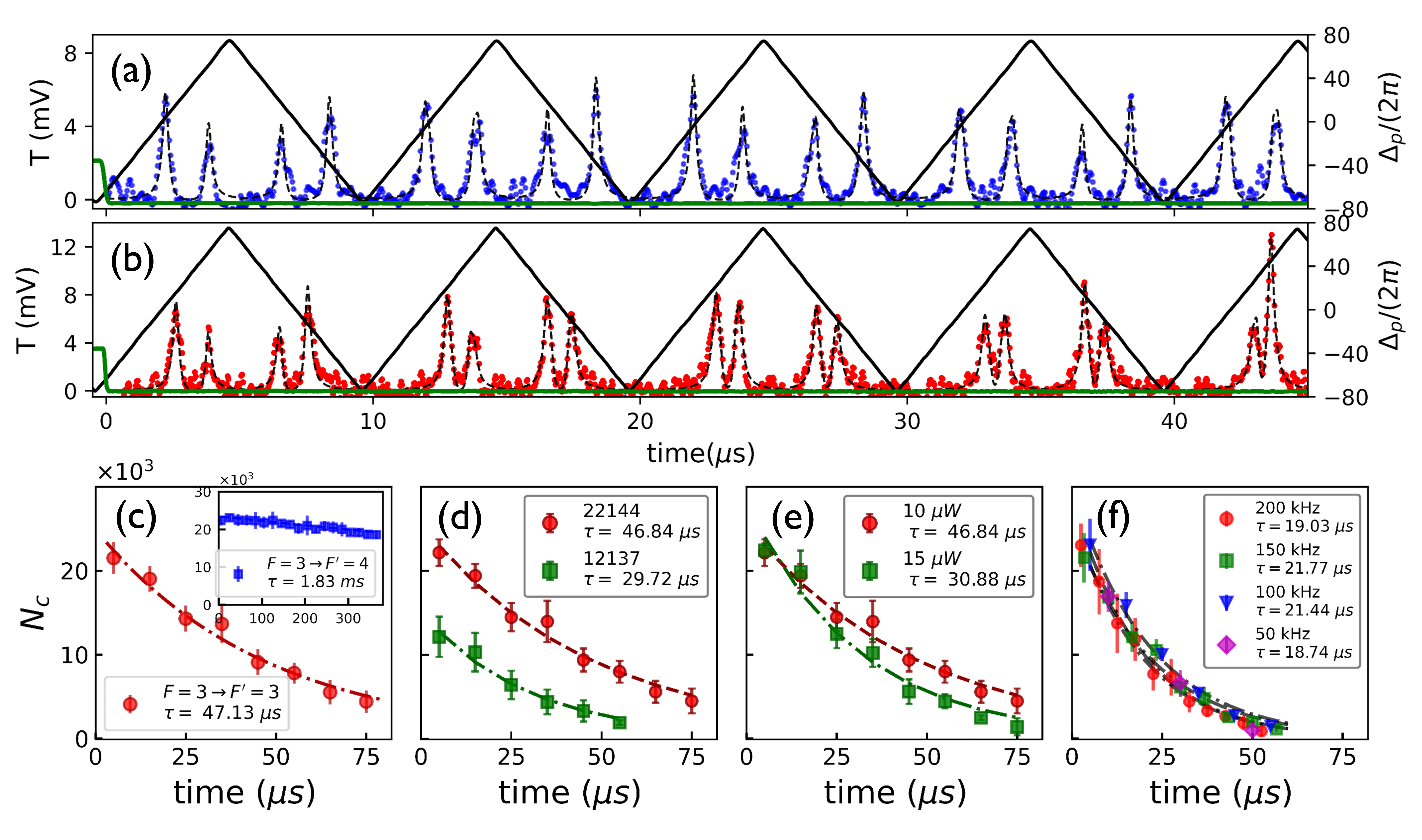}
	\caption{(a) Time evolution of VRS values for closed transition (blue) and (b) open transition (red). Here, the probe laser is locked to the corresponding atomic transition and scanned using AOM (black line). The green curve shows the trigger used for timing the switching off of cooling and repumper. (c) Time evolution of $N_c$, calculated from VRS using VRS$=2g_0\sqrt{N_c}$ for the open transition. The inset is the case for a closed transition. The dot-dashed line represents a single exponential fit in all cases. The time evolution of open transition (d) for different initial atom number $(N_c^0)$, (e) probe power, and (f) scan rate. }
	\label{fig2}
\end{figure*}

Fluorescence detection of atoms with an open transition (Fig.~\ref{fig1} (a)) requires measurements to be made on a timescale comparable to the total decay rate of the excited state. Fig.~\ref{fig1} (b) shows the evolution of the population of each state when probed across $|1\rangle\leftrightarrow|3\rangle$ transition. This numerical calculation is done using QuTip~\cite{qutip}. The system reaches a steady state in $\sim 200$ ns when the entire atomic ensemble ends up in the dark state, $|2\rangle$. Such a measurement is experimentally challenging even for a large ensemble of atoms, as shown later.

Using an optical cavity can significantly enhance the detection process for the same system. When an ensemble of atom $(N_c)$ is within the cavity mode volume, the collective coupling of atoms to the cavity results in vacuum Rabi splitting (VRS)~\cite{normal-mode1,normal-mode2,normal-mode3,jaynes-cumming}, which causes the cavity transmission to be at $\pm g_0\sqrt{N_c}$ (fig.~\ref{fig2} (c)) from atomic transition~\cite{tavis-cummings,dan-stamper1,sourav-ion,gokul-pa}, where $g_0$ is the single atom-cavity coupling strength. Due to the detuning of the probe laser from the $|1\rangle\leftrightarrow|3\rangle$ transition, the probability of decay to the state $|2\rangle$ is considerably reduced. As seen in Fig.~\ref{fig1} (d), the state evolution to the final state $|2\rangle$ happens over three orders of magnitude slower compared to the fluorescence case (Fig~\ref{fig1} (b)). For this numerical calculation, the density matrix for the coupled atom-cavity system is solved adiabatically (see eq.\ref{eq2}) with a low finesse cavity ($F\sim 330$) and atom number, $N_c = 20000$. In addition to the larger interrogation time, the cavity transmission directly measures the population in the ground state $(N_c)$ with a high signal-to-noise ratio (SNR). This provides a novel way to measure systems with open transitions compared to fluorescence detection and is readily adaptable for detecting an ensemble of molecules where the loss rates are more significant.

\section{\label{setup} Experimental setup \& Results}

The experimental setup consists of a cold cloud of ${}^{85}$Rb atoms trapped in a magneto-optical trap (MOT), which is co-centered with a low finesse ($F \sim 330$) FP cavity mounted inside a vacuum chamber. The MOT is formed by six independent cooling beams of diameter $\sim 10$mm, and the atoms are loaded from vapor in a controlled manner. The cooling beam is $-13$ MHz detuned from $F=3 \leftrightarrow F’=4$ $(3\leftrightarrow 4')$ transition, and the repumper beam is locked on $F=2 \leftrightarrow F’=3$ $(2\leftrightarrow 3')$ transition.  For typical operation conditions, the peak density of MOT is $\sim 2\times10^{10}$ cm${}^{-3}$ and FWHM $\sim 200 \mu$m for a gradient field of $22$ G/cm. For the measurements involving closed and open transitions, the cavity is locked to $3 \leftrightarrow 4'$ and $F=3\leftrightarrow F=3'$ $(3\leftrightarrow3')$ respectively ~\cite{lock}.

Figure~\ref{fig2} (a) shows the time evolution of VRS in a closed two-level system ($3\leftrightarrow 4'$, blue), and fig.~\ref{fig2} (b) shows that of open transition ($3\leftrightarrow 3'$, red) after turning off the cooling and repumper lasers. The dashed line shows the theoretical fit of intra-cavity intensity for a two-level system coupled to a cavity (see SM eq.S$1$). The difference in the VRS peak height is due to power variation across the AOM scan. In addition, the loss of atoms within the scan results in small changes in frequency and peak height in VRS for open transitions. This asymmetry does not affect the calculated value of $N_c$ in the experiment (see SM section S3).

Fig.~\ref {fig2}. (c) shows the evolution of $N_c$  in an open transition. Here, $N_c$ is calculated from VRS using the relation VRS $=2 g_0 \sqrt{N_c}$, where $g_0 = -\mu_{ge}\sqrt{\omega_{cv}/( 2\hbar\epsilon_0 V_c)} \approx 0.13$ MHz for $3 \leftrightarrow 3'$ and $\approx 0.2$ MHz for $3 \leftrightarrow 4'$. Here, $\omega_{cv}$ and $V_c$ are the cavity resonance frequency and cavity mode volume, respectively. The data points represent the value of $N_c$ after each scan. The decrease in the $N_c$ can be attributed to the loss of atoms to $F=2$ due to a weak non-resonant cavity probe. The loss rate $(\tau)$ of the transition is determined from a single exponential fit (dash-doted line), and for the open transition, is $\tau_o \approx 47$ $\mu $s. The inset in fig.~\ref{fig2}(c) shows the time evolution for the closed transition with $\tau_c \approx 1.8$ ms, which corresponds to the time scale of MOT expansion~\cite{tridib-temp}.

Fig.~\ref {fig2}(d), (e), and (f) show the dependences of decay time on $N_c$, probe power, and scan rate. From the figure~\ref{fig2}(d), the time constant increases with an increase in $N_c$. This increase in decay time is because, for large $N_c$, the cavity VRS is further away from atomic resonance, causing a smaller fraction of atoms to decay to the state $F=2$. The increase in the time constant in fig.~\ref{fig2}(e) is attributed to decreases in the probability of transition to the state $F=2$ with a reduction in the probe power. Here, the probe power mentioned is the total input power to the cavity, which is much larger than the effective power coupled to the cavity mode. In addition, measurements with different scan rates of the probe laser did not affect the decay time scale in the current experiment (fig.~\ref{fig2}(f)). These show that the loss rate due to the measurements depends only on the probe power and detuning of the VRS due to atom number. As a result, a near non-destructive detection of atoms/molecules with open transitions can be achieved by probing a large ensemble coupled to the cavity with a very weak probe.

\section{\label{sec:mod} Model \& Discussion}

The theoretical model for a simplified three-level atomic system coupled to a cavity ~\cite{rahul-molecule} is extended for studying atoms with the open transition, as shown in fig.~\ref{fig1}(c). Consider the decay rate from excited state $(|3 \rangle )$ to ground $(|1\rangle )$ and dark state $(|2\rangle)$ as $\Gamma_{31}$ and $\Gamma_{32}$ respectively. For N three-level system inside the cavity mode, the Hamiltonian under rotating wave approximation is given by 
\begin{equation}
    H = \hbar \sum_{k=1}^{N} \left[ -\Delta_{pa}\sigma_{33}^k + g_j \left( \hat{a} \sigma_{31}^{k}+ \hat{a}^{\dag}\sigma_{13}^k \right) \right]
\end{equation}
where $a (a^{\dag})$ is the photon annihilation(creation) operator, $\sigma_{ij} = |i\rangle\langle j|$ is the atomic operator, $\Delta_{pa} = \omega_p - \omega_a$ is the probe detuning from atomic transition, $g_j=g_0\sqrt{N_c}$ is the effective atom-cavity coupling constant.

\begin{figure}[t!]
    \centering
    \includegraphics[width=0.47\textwidth]{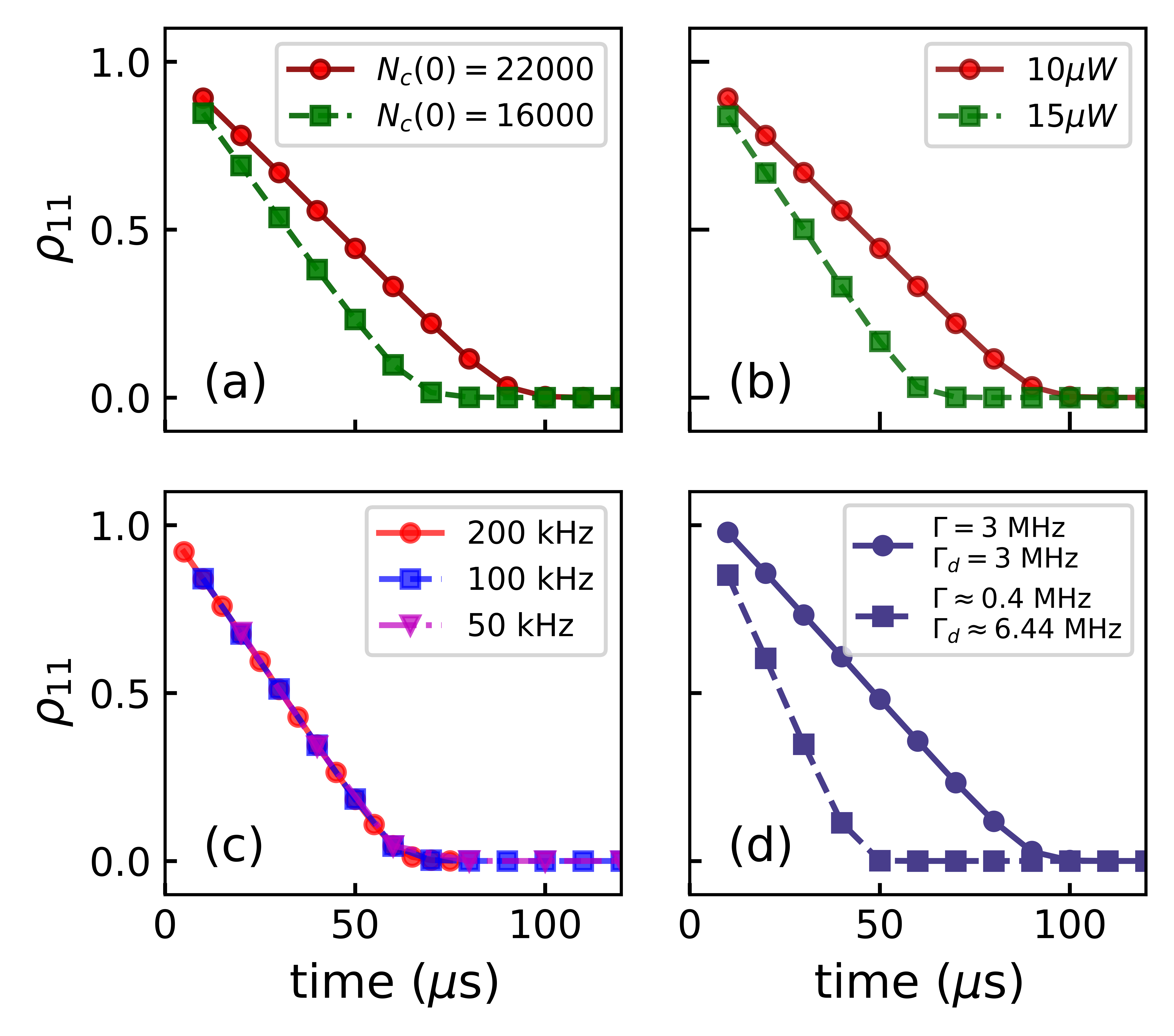}
    \caption{Evolution of ground state population, $\rho_{11}$ as a function of time (a) for different $N_c^0$, (b) different probe power, and (c) different scan rates. (d) A comparison of the loss rate of population in atomic transition (fig.~\ref{fig1} (a), solid line with circle) to molecular transition (dashed line with square) in Rb${}_2$. Here, the relevant molecular transition used are $(B{}^1 \Pi_u (\nu_e = 1, J_e=1) \leftrightarrow X{}^1\Sigma_{g}^{+} (\nu_g=0, J_g =0))$.~\cite{rahul-molecule}}
    \label{fig3}
\end{figure}

Assuming the cavity field is given by a classical coherent field $|\alpha \rangle$, the system's time evolution is given by the following coupled differential equations of density matrix $(\rho)$ and $\alpha$ ~\cite{Scully_Zubairy_1997,rahul-molecule}.
\begin{equation}\label{eq2}
    \begin{split}
        &\frac{d\alpha}{dt} = -(\kappa_t-i\Delta_{pc})\alpha -ig_0 N_c \rho_{13} - \eta \\
        &\frac{d\rho_{33}}{dt} = -\Gamma_t\rho_{33} + ig_0 \left(\alpha^*\rho_{13}-\alpha\rho_{31}\right) \\
        &\frac{d\rho_{13}}{dt} = -\left(\Gamma_t/2 -i\Delta_{pa}\right) \rho_{13} + ig_0\alpha\left(\rho_{33}-\rho_{11}\right) \\
        &\frac{d\rho_{11}}{dt} = \Gamma_{31}\rho_{33} - ig_0\left(\alpha^*\rho_{13}-\alpha\rho_{31}\right) \\
        &\frac{d\rho_{22}}{dt} = \Gamma_{32}\rho_{33}
    \end{split}
\end{equation}
where $\Gamma_t = \Gamma_{31}+\Gamma_{32}$ and $\eta$ is the rate at which the input field is transmitted into the cavity. This set of coupled equations can be solved numerically to calculate the time evolution of the population of the atom-cavity system. For a scan rate of $100$ kHz and scan amplitude of $\omega_a \pm 80$ MHz, the equation is solved adiabatically to calculate the time evolution of cavity field $(|\alpha|^2)$ and ground state population $(\rho_{11})$. Fig.~\ref {fig3}(a), (b), and (c) shows the time evolution of the ground state population, which is a direct measure of $N_c$, from the model for different $N_c$, probe power, and scan rates. From the model, the loss rate of the $\rho_{11}$ decreases with an increase in $N_c$, a decrease in power, and is independent of the scan rate as observed in the experiment (fig.\ref{fig2}).

\begin{figure*}[!t]
   \centering
   \includegraphics[width=0.87\textwidth]{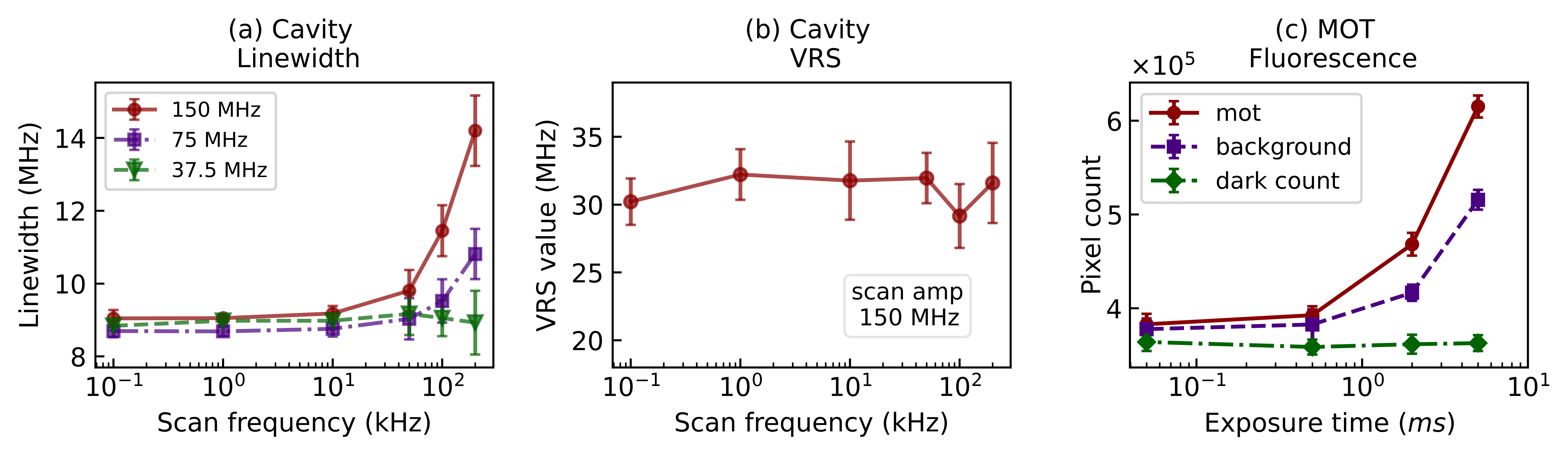}
    \caption{(a) Empty cavity linewidth measured at different scan frequency and frequency ranges. Here, the red curve is for a scan range of $150$ MHz, purple is for $75$ MHz, and green is for $37.5$ MHz (b) Cavity VRS measurement for different scan frequencies for a scan range of $150$ MHz. Each data point for (a) and (b) is the mean of 20 measurements, and the error bar is the standard deviation. (c) MOT fluorescence measurement for different exposure times. Each data point represents the mean pixel count for a $288 \times 300$ image, and the error bar is the $1\sigma$ error. Here, the red curve is the MOT count, purple is the background count, and green is the dark count. This measurement is performed with a total MOT number $\sim 5\times 10^5$ using the imaging setup~\cite{tridib-temp}. }
    \label{fig4}
\end{figure*}

This model does not include additional losses and other state-dependent effects. The time evolution of closed transition was measured for different $N_c$, which shows that the effects of additional state-dependent losses in our system are much lower than the open transition rates (see SM section S4). As a result, the effect of these losses will be minimal in our system. Despite not including additional losses, the model qualitatively explains all the experimental observations in fig~\ref{fig2}.

Numerical calculations were done for the case of Rb${}_2$ to see the possibility of extending this detection method to molecules with the current setup. Fig.~\ref{fig3} (d) shows the expected population evolution of Rb${}_2$ with rates $\Gamma = 401$ kHz $(B{}^1 \Pi_u (\nu_e = 1, J_e=1) \leftrightarrow X{}^1\Sigma_{g}^{+} (\nu_g=0, J_g =0))$ and $\Gamma_d \approx 6.44$ MHz (total decay to all other states)~\cite{rb2_1,rb2_2,rb2_3}. As seen in the figure, observing the population evolution requires an initial molecule of $\sim 20000$ in the ground state, which is experimentally challenging. Designing the cavity specific to the molecular scheme makes this detection feasible with $2000$ molecules, as shown in ~\cite{rahul-molecule}.

As discussed in fig~\ref{fig1}(b), direct fluorescence detection of open transition is experimentally challenging. A series of experiments were performed to understand the limit of fluorescence detection in the current experimental setup and compare this with the cavity-based detection scheme (details SM section S7). Empty cavity linewidth and cavity VRS for the closed transition $(3\leftrightarrow 4')$ were measured at different probe scan frequencies. Fig.~\ref{fig4} (a) shows that the linewidth remains constant for low scan frequencies and increases for large scan ranges at higher frequencies. This increase in the cavity linewidth happens when the effective time required to sweep across cavity resonance is  $\lesssim 1/\kappa_t$, which is the time required for the system to reach a steady state. Irrespective of the increase in linewidth, the VRS values remain constant for all scan frequencies and ranges (fig.~\ref{fig4}(b)) and are only limited by the bandwidth of the scan. This shows that VRS is a reliable measure for studying state-dependent dynamics, even at a very short time scale. In addition, MOT and background images were measured at different exposure times using the current imaging setup~\cite{tridib-temp}. As seen from fig.~\ref{fig4}(c), the SNR is very low for an exposure time of $\lesssim 1$ ms, making fast measurements difficult using fluorescence detection. A high numerical aperture imaging setup can increase the total fluorescence signal collected, thereby increasing the SNR. While implementing a high NA setup itself is challenging, especially for hybrid trap systems, detecting a low atom number makes the experimental realization of free space fluorescence detection of atoms with open transition extremely difficult, thus illustrating the advantages of cavity-based detection. An alternate cavity fluorescence-based measurement has been reported for fast measurements with single atom~\cite{dan-stamper2}.

\section{\label{sec:con} Conclusion}
In this paper, we have experimentally demonstrated the decay rate in an open transition using a collective strong coupling of atoms to an FP cavity. This rate depends on probe power and the initial number of atoms in the cavity mode volume. These are compared with that of a closed two-level system. Further experimental results are explained using a simple three-level model. Implementing this technique as a nearly non-destructive detection for molecular states can provide an alternate and more straightforward way for state detection and studying interactions.

Detecting fluorescence from 20000 atoms in the time scale of a few microseconds is experimentally challenging. This requires high bandwidth, low dark count detectors, and a high collection efficiency optical imaging system. Having a cavity converts this detection from intensity measurement to frequency measurement. In addition, measuring cavity transmission ensures high SNR due to minimal background. This allows for a much easier experimental scheme for such fast, low atom number measurements in an open transition and makes such measurements useful for high-fidelity, background-free state detection. Another advantage of cavity detection is that optical elements are not required to be placed in close proximity as in fluorescence measurements. The VRS scheme described in Sawant \textit{et al.} ~\cite{rahul-molecule} is fully validated by the present experiment.

\begin{acknowledgments}
We thank Meena M S for technical help. Funding from Department of Science and Technology and Ministry of Electronics and Information Technology (MeitY), Government of India, under a Centre for Excellence in Quantum Technologies grant with Ref. No. 4(7)/2020-ITEA.
\end{acknowledgments}
\bibliography{main}

\end{document}


\title{Supplemental material: Detection of radiatively open systems using an optical cavity }

\author{V. I. Gokul${}^1$, Arun Bahuleyan${}^1$, Raghuveer Singh Yadav${}^1$,  S. P. Dinesh${}^1$, V. R. Thakar${}^1$,\\ Rahul Sawant${}^2$ \& S. A. Rangwala${}^1$}
\affiliation{%
 ${}^1$Raman Research Institute, C.V. Raman Avenue, Sadashivanagar, Bangalore 560080, India \\
 ${}^2$ I-Hub Quantum Technology Foundation, Indian Institute of Science Education and Research Pune, Dr. Homi Bhabha Road, Pashan, Pune 411008 India.
}%
\maketitle
\section{Experimental Method}
For measuring the time evolution of VRS, the cavity is locked to the desired atomic transition~\cite{lock} using a weak probe, referenced to saturated absorption spectroscopy. This ensures that there is no drift in the cavity frequency throughout the experiment. Once the cavity is locked, the probe laser is also locked to the same transition. Two double-pass AOMs are used in the probe path, one to scan the frequency of the locked laser and the other to compensate for the frequency shift. The AOM is scanned from a function generator with a maximum triangular scan frequency of $200$ kHz. Once the MOT is loaded to number saturation, the cooling and repumper beams are turned off using an AOM-based switch for a time duration of $2$ ms. This sequence is repeated at a $10$ Hz repetition rate. The efficiency of the switch is independently verified by looking at the power output of cooling and repumper using a photodiode.  The probe AOM is scanned at a rate of $100$ kHz and the evolution of VRS is measured using a PMT with a preamp (Hamamatsu-C9999), which has a $10$ MHz bandwidth. The experiment is repeated for different probe powers and MOT atom numbers.

\section{Intra-cavity intensity for a two-level atom-cavity system}
To calculate $N_c$ from the VRS measurement, a theoretical fit for intracavity intensity for a two-level system coupled to a cavity given by the Jaynes-Cummings model~\cite{jaynes-cumming} is used. This is given by
\begin{equation}\label{intra-cav}
    |\alpha|^2 = \frac{|\eta|^2(\frac{\Gamma^2}{4}+\Delta_{pa}^2)}{(\frac{\kappa_t\Gamma}{2}-\Delta_{pa}^2+g_t^2)^2+(\kappa_t+\frac{\Gamma}{2})^2\Delta_{pa}^2}
\end{equation}
 
Here, $\eta$ is the probe power injected into the cavity, and $\Delta_{pa}$ is the probe detuning from atomic transition. The final form of the fit used to calculate VRS value is $|\alpha_1|^2 + |\alpha_2|^2$~\cite{gokul-pa}. For the fit the values used for different quanities are $\Gamma = 6.06$ MHz, $2\kappa = 9$ MHz, $g_t = g_0\sqrt{N_c}$, where $g_0 = 0.2$ MHz for $F=3\rightarrow F'=4$ and $g_0=0.131$ MHz for $F=3\rightarrow F'=3$. From the fit, the value of the number of atoms in the cavity mode volume, $N_c$, is extracted. The experiment is repeated multiple times to get the mean and standard deviation. 
\begin{figure}[!h]
    \centering
    \includegraphics[width=0.67\textwidth]{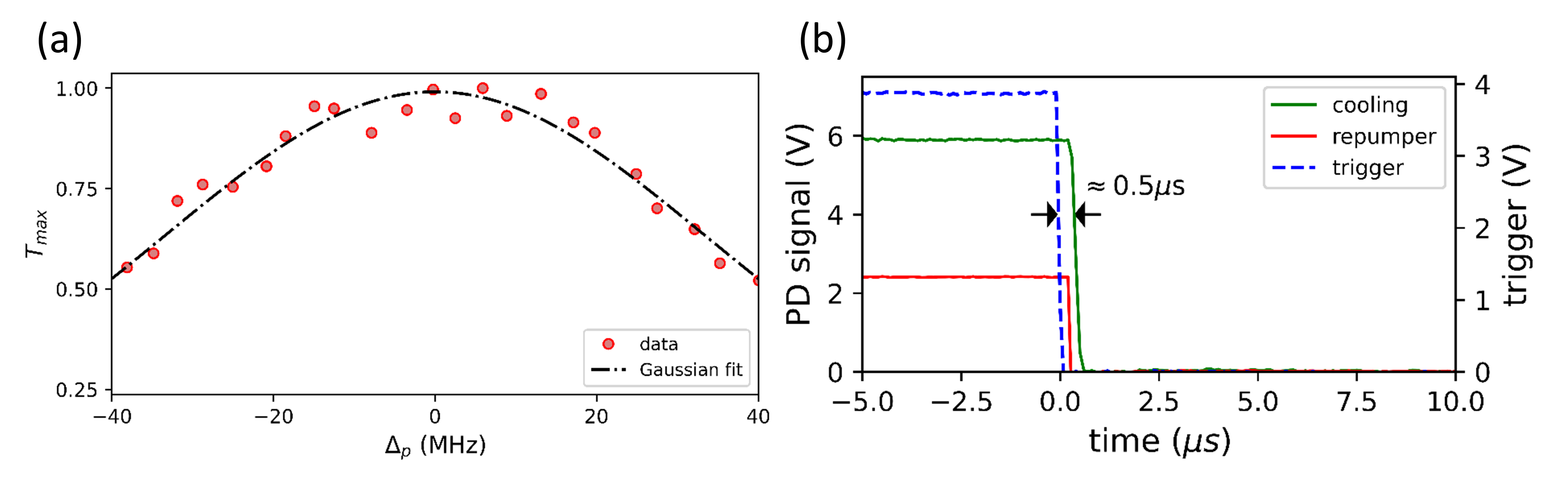}
    \caption{(a) AOM power variation. Here, the maximum cavity transmission is plotted for different atom-cavity detuning. Each point represents the experimental value, and the dashed line is a Gaussian fit. (b) AOM switch for cooling and repumper. Here, the red and green curve represents the photodiode value of repumper and cooling when an AOM-based switch is implemented. The dashed blue curve is the trigger pulse used for switching}
    \label{Aom}
\end{figure}

\section{AOM power variation and switching}

The probe laser is locked to the SAS signal, and one of the double-pass AOM's drive frequencies is varied. The figure.~\ref{Aom} (a) shows the variation in the maximum transmission as a function of atom-cavity detuning. This variation in the power is due to the change in efficiency of AOM for different RF drive frequencies. The dashed line in the figure shows a Gaussian fit to the data. This functional form of power variation is used in the model to consider the effects of power variation. Figure.~\ref{Aom} (b) shows the photo-diode signal showing the switching off cooling and repumper during the experiment. As seen from the figure the switch for cooling and repumper is almost instantaneous with high fidelity. A small delay of $\approx 0.5 \mu$s between the trigger pulse and the actual switching of AOMs does not affect the experiment. This ensures that the system has no cooling or repumper light during the experiment.

\section{Additional state-dependent losses}
\begin{figure}[!h]
    \centering
    \includegraphics[width=0.67\textwidth]{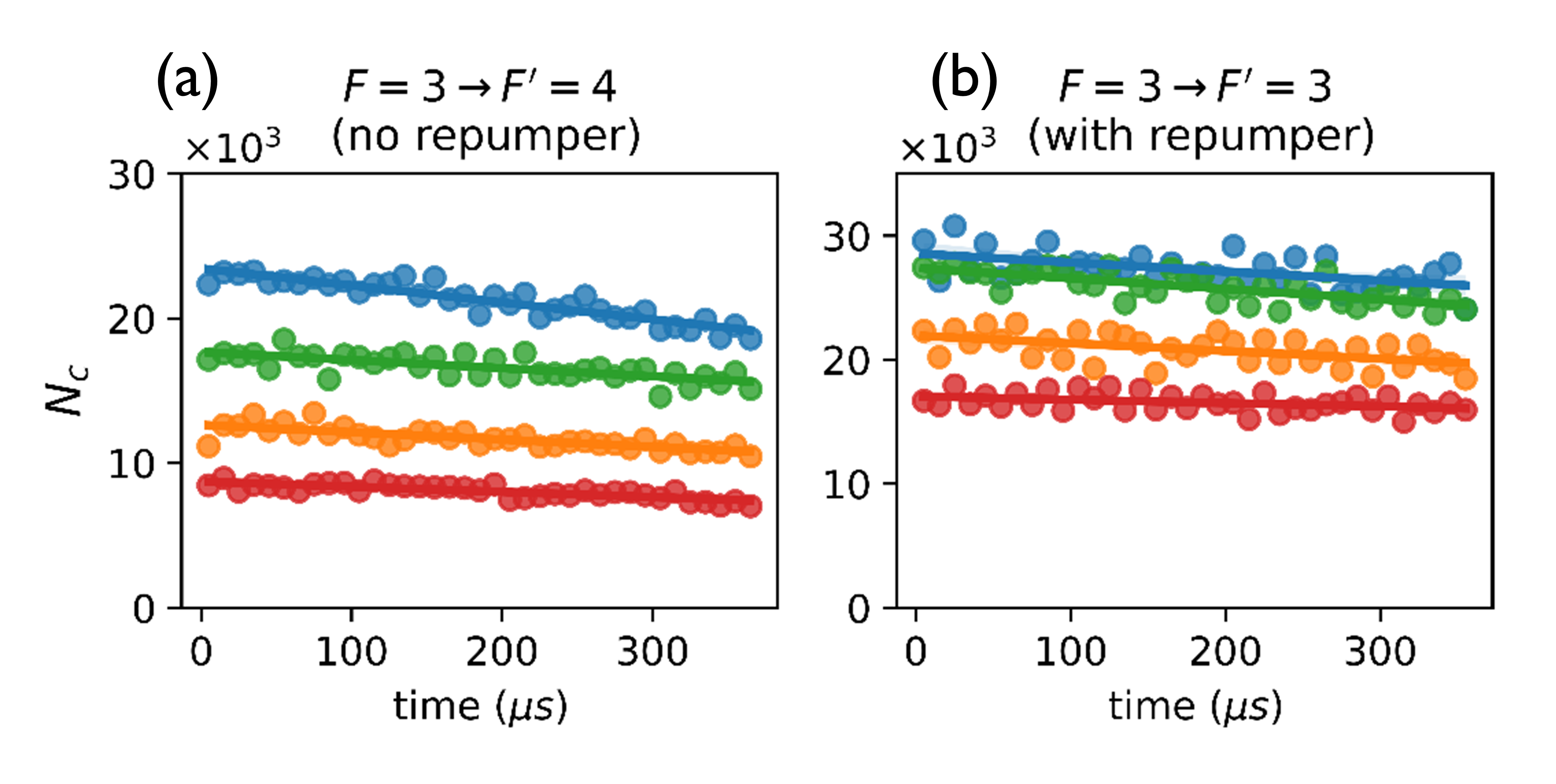}
    \caption{Time evolution of $N_c$ for (a) closed two-level system coupled to cavity (b) open transition with repumping laser. Here, blue, green, orange, and red represent an initial atom number of $\sim$ $22000$, $17000$, $11000$, and $8500$, respectively. All the data point is the mean, and the error bar is the standard deviation of 5 measurements.}
    \label{add-loss}
\end{figure}
To calculate additional state-dependent losses in the system, measurements were done with closed transition $(3\rightarrow4')$ for different $N_c$. From fig.~\ref{add-loss} (a), the loss rate increases with $N_c$, which shows that there are additional losses in the system that directly depend on $N_c$. Since this loss occurs over a much longer timescale than the open transition, it does not affect the rates of open transition.  The same measurement was also repeated in an open transition with an additional repumping laser $(F=2\rightarrow F'=3)$ that plugs the decay channel, making it a case similar to a closed transition. Figure~\ref{add-loss} (b) shows the corresponding loss rate with $N_c$ for an optically plugged open transition. As seen from the figure, the increase in loss rate with $N_c$ is considerably less compared to the closed transition. This shows that most of the losses result in an additional decay path to the dark state $(F=2)$.

\section{3-level open transition model}\label{intra-cav}
\begin{figure}[!h]
   \centering
   \includegraphics[width=0.67\textwidth]{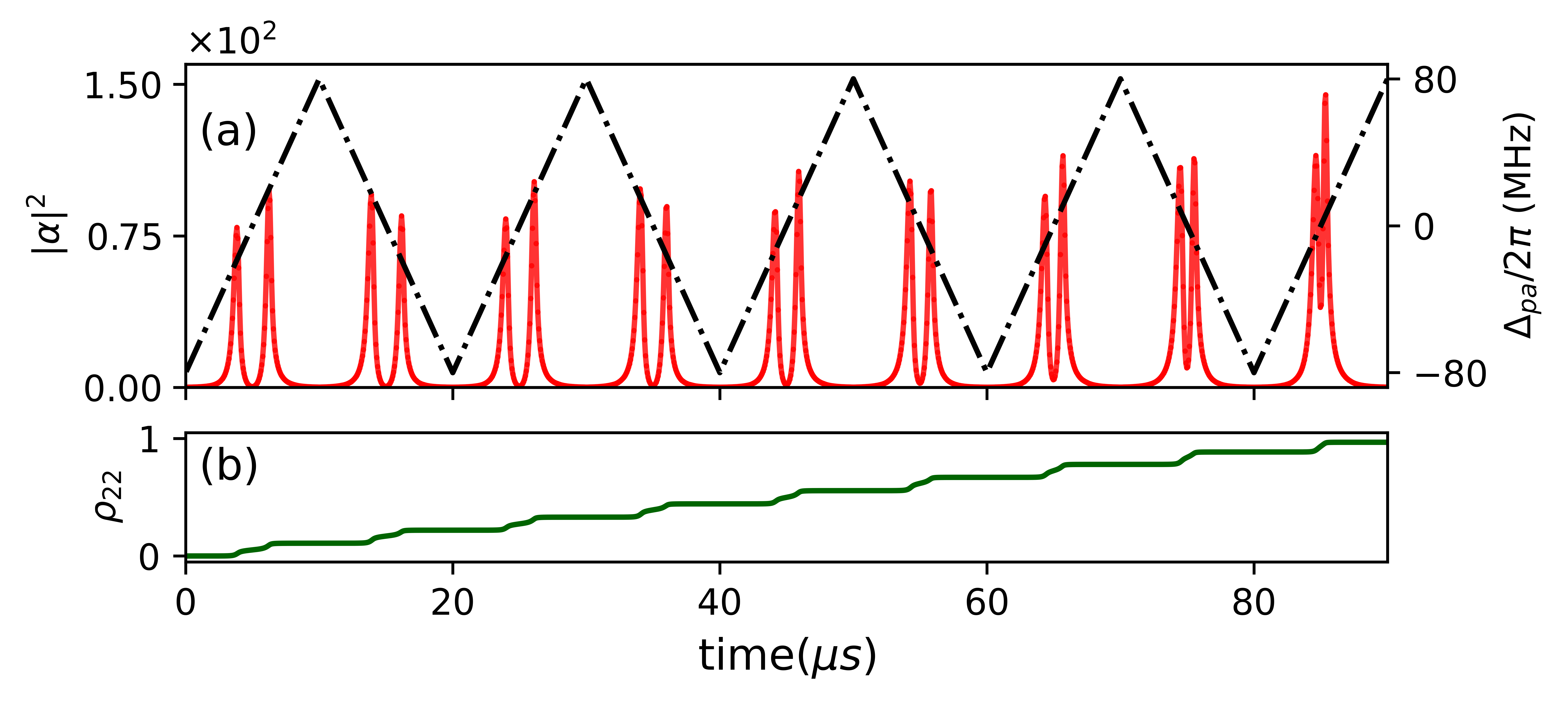}
   \caption{(a) Time evolution of VRS calculated numerically using the model. Here, $p_{out}$ is the cavity output power calculated from the intra-cavity intensity $|\alpha|^2$, dashed black curve is the AOM scan. (b) The time evolution of population $(\rho_{22})$ in dark state $(|2\rangle)$.}
   \label{model}
\end{figure}

The set of coupled equations (eq.2) for N three-level atoms interaction with cavity field can be solved numerically to calculate the time evolution of cavity field $(|\alpha|^2)$ and ground state population$(\rho_{11})$. The cavity output power is calculated from the intracavity intensity using~\cite{,rahul-molecule}
\begin{subequations}
\begin{align}
   E_{cv} &= \sqrt{\frac{\hbar\omega_{c}}{2\epsilon_0 V_c}}|\alpha|^2 \\
   I_{out} &= \frac{1}{2} c\epsilon_0 |E_{cv}|^2 \times T_2 \\
   p_{out} &= I_{out}\times \text{A}
\end{align}
\end{subequations}
where $E_{c}$ is the electric field inside the cavity, $T_2$ is the transmitivity of output mirror, and $A = 1/2 \pi w_0^2$ is the area of the cavity mode. The time for one scan $T=10 \mu s$ is divided into 300 steps of $\delta t = 10/300 \approx 0.033 \mu s$. The system is evolved for the time-period $\delta t$ with appropriate detuning, $\Delta_{pa}$. By the end of the scan total population lost to the dark state is calculated. This process is repeated for multiple scans, and the result is obtained.

The fig.~\ref{model} shows the time evolution of cavity output by numerically solving the coupled equation with current system parameters. As seen from the figure after each scan, there is a VRS reduction because of the population loss to the dark state $(|2\rangle)$. The population to the dark state $(\rho_{22})$ is plotted in fig.~\ref{model}, which shows an increase after each scan. This simple model explains most of the experimental results qualitatively.

\section{Signal processing and noise reduction}
\begin{figure}[!h]
    \centering
    \includegraphics[width=0.47\textwidth]{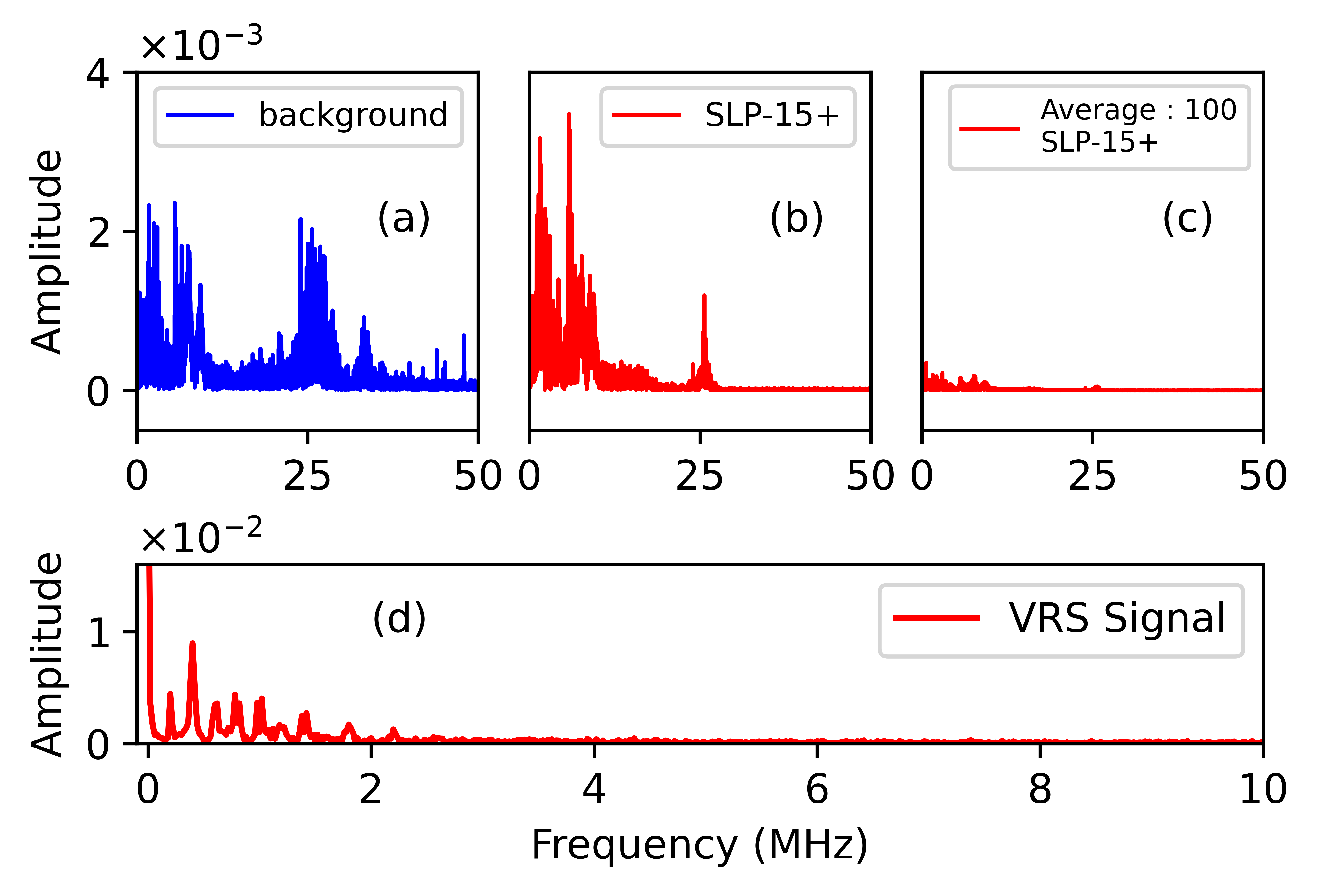}
    \caption{Fast Fourier transform (FFT) of background noise (a) without filter, (b) with low pass filter (SLP-15+), (c) with a low pass filter and an averaging of 100 measurements, and (d) FFT of VRS signal}
    \label{noise}
\end{figure}
A higher signal-to-noise ratio (SNR) is important for all measurements. The background of cavity transmission being nearly zero makes cavity-based detection a high SNR measurement. To further improve the SNR for fast measurement (with scan rate up to $200$ kHZ), a low pass filter of 15 MHz bandwidth (Minicircuit SLP-15+) is used. Figure ~\ref{noise} shows the FFT of the background with and without a low-pass filter. Further, an average of nearly 100 repeated measurements reduces the background noise. As seen from the figure~\ref{noise}, this results in an SNR of $\sim 50$.

\section{Characterization of cavity parameters for fast measurement}
\begin{figure}[!h]
    \centering
    \includegraphics[width=0.67\textwidth]{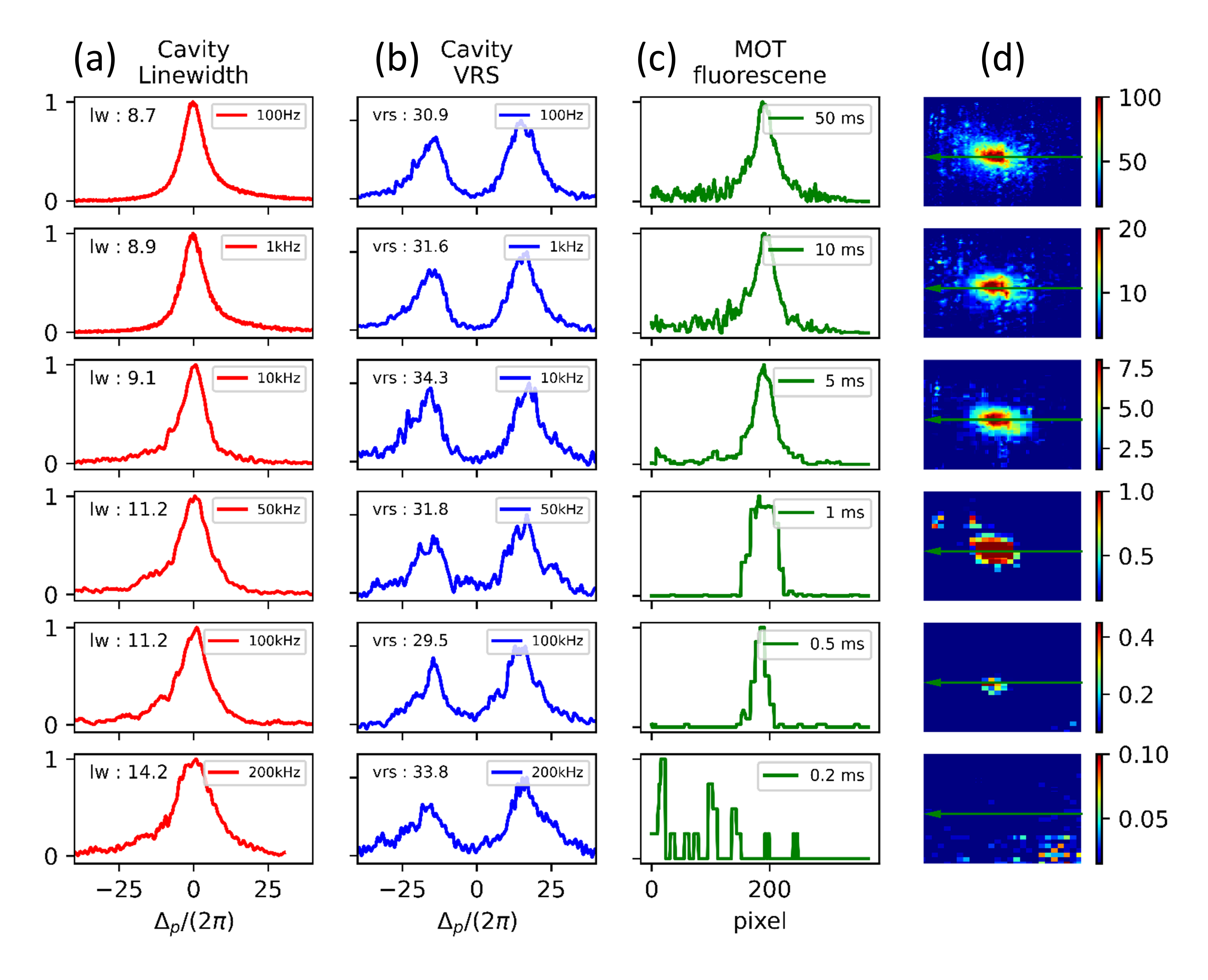}
    \caption{(a) Empty cavity linewidth measurement and (b) VRS measurements for different probe AOM scan frequencies with a scan range of $150$ MHz. (c) MOT intensity profile for different exposure times (d) Background subtracted MOT density profile at different exposure times. All the measurements were done for a closed transition and is an average of 100 measurements.}
    \label{scan_char}
\end{figure}

An explicit characterization has been done for different scan frequencies and scan ranges to measure the effect of fast measurement on the empty cavity linewidth and VRS. For linewidth measurement, the cavity is locked to $F=3\rightarrow F'=4$, and a laser locked to the same transition is used to probe the system. The empty cavity linewidth for different scan frequencies for a scan range of $\sim 150$ MHz is given in fig.~\ref{scan_char} (a). This data is fitted with a Lorenztian function to extract the linewidth, $\kappa$. As seen from the figure~\ref{scan_char} (a), the linewidth values increase with scan frequency. This increase in the cavity linewidth for a larger scan frequency is because the time required to sweep across cavity resonance is less than that required for the cavity to reach a steady state. This results in a larger empty cavity linewidth from its steady state value. Figure.~\ref{scan_char} (b) shows the VRS measurement for different scan frequencies with a probe scan range corresponding to $150$ MHz. The data is fitted with a theoretical model for intracavity intensity (eq.\ref{intra-cav}) to extract the value of VRS. As seen (fig 4 (b)), the value of VRS remains constant within the error bar for all the scan frequencies. This shows that the measurements that rely on cavity linewidth for state detection are not well-suited for fast measurements without incorporating these additional linewidth changes. In contrast, VRS is a reliable measure for studying dynamics at a shorter time scale. To compare these measurements with fluorescence detection in the current experimental setup~\cite{tridib-temp}, MOT and background images were taken at different exposure times. Fig.~\ref{scan_char} (c) and (d) show the background subtracted image of MOT for various exposure times and corresponding intensity profiles. As seen from fig~\ref{scan_char}, the fluorescence measurement is unreliable for time scale $\sim 1$ms or lower with the current imaging setup.

\bibliography{supplemental}